\documentclass[journal=jacsat,manuscript=article]{achemso}
\setkeys{acs}{usetitle = true}
\usepackage[utf8]{inputenc}
\usepackage{color}

\usepackage[version=3]{mhchem} 
\usepackage{multirow}
\usepackage{natbib}

\usepackage[]{algorithm2e}

\author{Adriana Supady}
\email{supady@fhi-berlin.mpg.de}
\author{Volker Blum}
\affiliation[FHI Berlin]
{Fritz-Haber-Institut der MPG, Berlin, Germany}
\alsoaffiliation[Duke University]
{Department of Mechanical Engineering \& Materials Science, Duke University, Durham, U.S.A.}
\author{Carsten Baldauf}
\affiliation[FHI Berlin]
{Fritz-Haber-Institut der MPG, Berlin, Germany}
\email{baldauf@fhi-berlin.mpg.de}

\title{First-Principles Molecular Structure Search with a Genetic Algorithm}

\begin{document}




\section{Abstract}

The identification of low-energy conformers for a given molecule is a fundamental problem in computational chemistry and cheminformatics. We assess here a conformer search that employs a genetic algorithm for sampling the low-energy segment of the conformation space of molecules. The algorithm is designed to work with first-principles methods, facilitated by the incorporation of local optimization and blacklisting conformers to prevent repeated evaluations of very similar solutions. The aim of the search is not only to find the global minimum, but to predict all conformers within an energy window above the global minimum. The performance of the search strategy is: (i) evaluated for a reference data set extracted from a database with amino acid dipeptide conformers obtained by an extensive combined force field and first-principles search and (ii) compared to the performance of a systematic search and a random conformer generator for the example of a drug-like ligand with 43 atoms, 8 rotatable bonds and 1 \textit{cis/trans} bond. 

\section{Introduction}

One of the fundamental problems in cheminformatics and computational chemistry is the identification of three-dimensional (3D) conformers that are energetically favourable and likely to be encountered in experiment at given external conditions \cite{Schwab2010}. Conventionally, these conformers are often characterized by specific, fixed sets of nuclear coordinates or ensembles thereof, and their potential energy is given by the electronic degrees of freedom in a Born-Oppenheimer picture of the chemical bond. A variety of conformations can be adopted by flexible organic molecules as the multi-dimensional potential-energy surface (PES) usually contains multiple local minima, with a global minimum among them. Only when the relevant conformers are known, one can predict and evaluate chemical and physical properties of the molecules (e.g. reactivity, catalytic activity, or optical properties). In many practical applications, the PES minima are taken as starting points to explore the free-energy surface (FES).
Generating conformers is an integral part of methods such as protein-ligand docking \cite{Kuntz1982, Jones1997, Morris1998, Meier2010} or 3D pharmacophore modeling \cite{Kristam2005}. The propensity to adopt a certain conformation strongly depends on the environment and possible interactions with other compounds. It has been shown that the bioactive conformation of drug-like molecules can be higher in energy than the respective global minimum \cite{Kirchmair2005} and that different 3D conformations may be induced by specific interactions with other molecules \cite{Agrafiotis2007}. Thus, it is crucial to focus not just on a single, global minimum of the PES, but instead to provide a good coverage of the accessible conformational space of a molecule yielding diverse low-energy conformers.
 
The exploration of a high-dimensional PES is challenging. A selection of popular sampling approaches utilized in conformer generation is summarized in Table~\ref{table:tbl_sampling}.
\begin{table}[h]
  \caption{Popular sampling approaches. Names of freely available programs are highlighted in boldface.}
  \label{table:tbl_sampling}
\noindent 

  \small

\begin{tabular}{|p{2.9cm}|p{7.5cm}|p{5.0cm}|}

 \hline
 Method & Description & Implemented, e.g., in \\

\hline
\hline
grid-based & based on grids of selected Cartesian or internal coordinates (e.g., grids of different torsional angle values of a molecule) & CAESAR\cite{Li2007}, \textbf{Open Babel}\cite{O'Boyle2011a}, \textbf{Confab}\cite{O'Boyle2011b}, MacroModel\cite{Mohamadi1990}, MOE\cite{moe}  \\
\hline
rule/knowledge - based & use known (e.g., from experiments) structural preferences of compounds  & \textbf{ALFA}\cite{Klett2014}, \textbf{CONFECT}\cite{Scharfer2013}, CORINA and ROTATE\cite{Sadowski1994, Renner2006}, \textbf{COSMOS}\cite{Andronico2011, Sadowski2013}, OMEGA\cite{Hawkins2010}\\ 
\hline
population-based metaheuristic & improve candidate solutions in a guided search & \textbf{Balloon}\cite{Vainio2007}, \textbf{Cyndi}\cite{Liu2009}\\
\hline 
distance geometry & based on a matrix with permitted distances between pairs of atoms & \textbf{RDKit} \cite{rdkit} \\ 
\hline
basin-hopping \cite{Wales1997}  / minima hopping \cite{Goedecker2004} & based on moves across the PES combined with local relaxation & \textbf{ASE} \cite{Bahn2002}, \textbf{GMIN}\cite{gmin}, \textbf{TINKER SCAN} \cite{tinker}  \\

\hline
  
  \end{tabular}
\end{table}
We focus specifically on genetic algorithms (GAs) \cite{Holland1975, Fogel1998, Goldberg1989} that belong to the family of evolutionary algorithms (EAs) that are frequently used for global structure optimization of chemical compounds \cite{Clark1996, Wales2003, Wales1999, Johnston2004, Nair1998, Damsbo2004, Carstensen2011, Carlotto2012, Jones1997, Morris1998,Hartke1993, Deaven1995, Hartke1999, Johnston2003, Blum2005, Schonborn2009, Sierka2010, Bhattacharya2013,Hartke2011, Heiles2013,Hart2005, Abraham2006, Oganov2006,Woodley2008,Tipton2013, Neiss2012, Brain2011}. GAs for chemical structure searches implement a 'survival of the fittest' concept and adopt evolutionary principles starting from a population of, most commonly, random solutions. GAs use the accumulated information to explore the most promising regions of the conformational space. With this, the number of unhelpful evaluations of physically implausible high-energy solutions can be reduced. Examples of GA-based structure prediction applications include: (i) conformational searches for molecules like of unbranched alkanes \cite{Nair1998} or polypeptide folding \cite{Damsbo2004}; (ii) molecular design \cite{Carstensen2011, Carlotto2012}; (iii) protein-ligand docking\cite{Jones1997, Morris1998}; cluster optimization \cite{Hartke1993, Deaven1995, Hartke1999, Johnston2003, Blum2005, Schonborn2009, Sierka2010, Bhattacharya2013,Hartke2011, Heiles2013}; (v) predictions of crystal structures\cite{Hart2005, Abraham2006, Oganov2006,Woodley2008}; (vi) structure and phase diagram predictions \cite{Tipton2013}. Further, Neiss and Schoos \cite{Neiss2012} proposed a GA including experimental information into the global search process by combining the energy with the experimental data in the objective function. Since GAs typically rely on internal, algorithmic parameters that control the efficiency of a search, a meta-GA for optimization of a GA search for conformer searches was proposed by Brain and Addicoat \cite{Brain2011}. 

Aside from the search algorithm itself, the choice of the mathematical model for the PES is critical to ensure results that reliably reflect the experimental reality. Among the available atomistic simulation approaches, "molecular mechanics" models, i.e., so-called force fields are especially fast from a computational point of view and therefore often employed. However, the resulting predictions depend on the initial parametrization of a particular force field and can lead to considerable rearrangements of the true PES for molecules that were not included in the parameterization procedure \cite{Baldauf2013, Rossi2014, Avgy-David2015}. On the other end of the spectrum of approaches, the PES can be faithfully represented based on the "first principles" of quantum mechanics. Indeed, benchmark quality approaches such as coupled cluster theory at the level of singles, doubles and perturbative triples (CCSD(T)) are almost completely trustworthy for closed-shell molecules, but still prohibitively expensive towards larger systems and/or large-scale screening of energies of many conformers. Density-functional theory (DFT) approximations are an attractive alternative to balance accuracy and computational cost. The choice of the approximation is critical when using first principles methods like DFT. It has been shown that it is necessary to incorporate dispersion effects for (bio)organic molecules and their complexes \cite{Wu2002, Sedlak2013, Tkatchenko2011, Baldauf2013}. The challenge of including long-range interactions has been met for example by the dispersion correction schemes described by Grimme \cite{Grimme2007, Grimme2010} or by Scheffler and Tkatchenko \cite{Tkatchenko2009, Tkatchenko2012, Ambrosetti2014}, but validating the DFT approximation employed is critical. In fact, subtle energy balances of competing conformers can require relatively high-level DFT approximations for reliable predictions \cite{Rossi2014, Schubert2015}.

The aim of our work is to develop and test an approach to sample the PES of small to medium sized (bio)organic molecules without relying on empirical force fields, utilizing instead electronic-structure methods for the entire search. With the molecular structure problem in mind, we define following requirements for the search strategy and implementation:
\begin{itemize}
\item Global search based on user-curated torsional degrees of freedom (bond rotations).
\item Local optimization based on full relaxation of Cartesian coordinates  and avoidance of recomputing too similar structures to ensure both efficient sampling and economic use of a computationally demanding energy function.
\item Design of the program in a way to use an external and easily exchangeable electronic structure code (in our case FHI-aims\cite{Blum2009, Havu2009}).
\item Simple input of molecules (composition and configuration) by means of SMILES codes\cite{Weininger1988}.
\item A robust and simple metaheuristic that ideally identifies the complete ensemble of low-energy conformers.
\item Freely available and with a flexible open-source license model.
\item Support for parallel architectures.
\end{itemize}
Based on these requirements, we present in this work a conformational search strategy based on a genetic algorithm. We provide a detailed description of our approach and a software implementation  Fafoom (Flexible algorithm for optimization of molecules) that is available under an open-source license (GNU Lesser General Public License\cite{gnuglgp}) for use by others. For simplicity, we abbreviate 'potential energy' with 'energy' and  'minima of the potential-energy surface' with 'energy minima'.

\section{Methods}

\noindent
In the following, we first motivate and explain assumptions that we met for handling 3D structures of molecules. Further, we outline the algorithm's implementation and describe its technical details. Finally we introduce a data set that we use as a reference for evaluating the performance of our implementation. Our work focuses on both the ability to reliably predict the global minimum and to provide a good conformational coverage with a computationally feasible approach. To achieve that, we formulate some specific algorithmic choices at the outset: (i) only sterically feasible conformations are accepted for local optimization; (ii) a geometry optimization to the next local minimum is performed for every generated conformation; (iii) an already evaluated conformation will not be evaluated again.

\subsection{Choice of coordinates}

In computational chemistry, at least two ways of representing a molecule's 3D structure are commonly used, either Cartesian or internal coordinates. The simplest internal coordinates are based on the 'Z-matrix coordinates', which include bond lengths, bond angles and dihedral angles (torsions) and can also be referred to as 'primitive internal coordinates'. These coordinates reflect the actual connectivity of the atoms and are well suited for representing curvilinear motions such as rotations around single bonds \cite{Schlegel2011}. Bond lengths and bond angles posses usually only one rigid minimum, i.e. the energy increases rapidly if these parameters deviate from equilibrium. In contrast, torsions can change in value by an appreciable amount without a dramatic change in energy. Similar to the work of Damsbo \textit{et al.} \cite{Damsbo2004}, we use Cartesian coordinates for the local geometry optimizations while internal coordinates, in this work only torsional degrees of freedom (TDOFs), i.e. freely rotatable bonds and, if present, \textit{cis/trans} bonds, are used for the global search. We consider only single, non-ring bonds between non-terminal atoms to be rotatable bonds after excluding bonds that are attached to methyl groups that carry three identical substituents. Further we allow for treating selected bonds in a \textit{cis/trans} mode, i.e. allowing only for two different relative positions of the substituents. In cases in which the substituents are oriented in the same direction we refer to it as to \textit{cis}, whereas, when the substituents are oriented in opposite directions, we refer to it as to \textit{trans}. 

\subsection{Handling of molecular structures}

\begin{figure}
\includegraphics{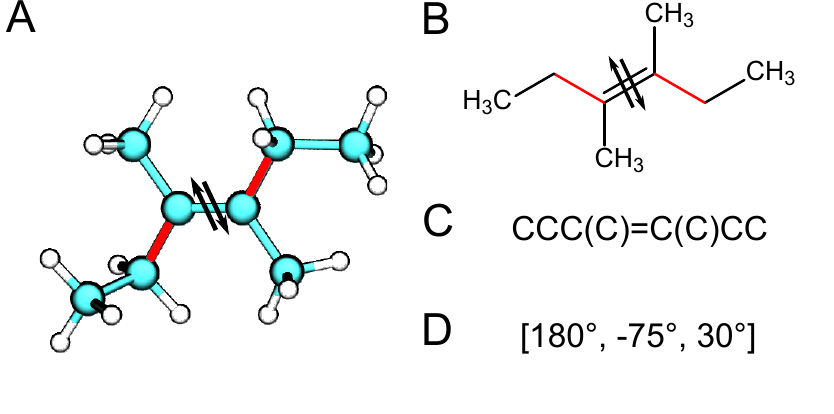}
 \caption{Different chemical representations of 3,4-dimethylhex-3-ene: (A) 3D structure  with rotatable bonds marked in red and the \textit{cis/trans} bond marked with double arrows. (B) 2D structure. (C) SMILES string. (D) vector representation of the molecule. The first value encodes the torsion angle value for the \textit{cis/trans} bond and the two remaining position store the torsion angle values of the rotatable bonds.}
  \label{fig_example}
\end{figure}

Figure~\ref{fig_example} shows different chemical representations of a molecule, here for the example of 3,4-dimethylhex-3-ene. Figure~\ref{fig_example}A and B depict the standard 3D and 2D representation of the compound together with marked \textit{cis/trans} and rotatable bonds. A SMILES (simplified molecular-input line-entry system) string is shown in Figure~\ref{fig_example}C. A SMILES  representation \cite{Weininger1988} of a chemical compound encodes the composition, connectivity, the bond order (single, double, triple), as well as stereochemical information in a one-line notation. Finally, a vector representation (Figure~\ref{fig_example}D) can be created if the locations of \textit{cis/trans} and rotatable bonds are known. The vector will store the corresponding torsion angle values. Our implementation takes as input a SMILES representation of a molecule, while vectors of angles are used to internally encode different structures in the genetic algorithm below.

\subsection{Frequently used terms}

Several terms need to be defined prior to describing the structure of the algorithm. In the following, the parameters of the search are highlighted in boldface. These parameters are input parameters to the algorithm and need to be defined in the input file.  

A \textit{sensible geometry} meets two constraints. First, the atoms are kept apart, i.e. none of the distances between non-bonded atoms can be shorter than a defined threshold (\textbf{distance\_cutoff\_1}, default=1.3 \r{A}). Secondly, it is fully connected, i.e. none of the distances between bonded atoms can be longer than a defined threshold 
(\textbf{distance\_cutoff\_2}, default=2.15 \r{A}). The attribute \textit{sensible} can be used further to describe any operation that outputs sensible geometries. 

The \textit{blacklist} stores all structures that: (i) were starting structures for the local optimization and (ii) resulted from local optimization, as they may have changed significantly during the optimization. In case of achiral molecules (\textbf{chiral}, default=False) also the corresponding mirror images are created and stored. 

A structure is \textit{unique} if none of the root-mean-square deviation (RMSD) values calculated for the structure paired with any of the structures stored in the blacklist is lower than a defined threshold (\textbf{rmsd\_cutoff\_uniq}, default=0.2 \r{A}). We consider only non-hydrogen atoms for the calculation of the RMSD.

\subsection{Basic outline of the search algorithm}
We implemented the genetic algorithm  (GA) using the Python language (version 2.7) and employ the RDKit library \cite{rdkit}. An overview is presented in Algorithm~\ref{alg:pseudocode}.

\SetStartEndCondition{ }{}{}%
\SetKwProg{Fn}{def}{\string:}{}
\SetKwFunction{Range}{range}
\SetKw{KwTo}{in}\SetKwFor{For}{for}{\string:}{}%
\SetKwIF{If}{ElseIf}{Else}{if}{:}{elif}{else:}{}%
\SetKwFor{While}{while}{:}{fintq}%
\AlgoDontDisplayBlockMarkers\SetAlgoNoEnd\SetAlgoNoLine%
\DontPrintSemicolon

\begin{algorithm}[H]
\setstretch{1.0}
\label{alg:pseudocode}

 \# initialization\;
 \While{i $<$ popsize}{
  x = random\_sensible\_geometry\;
 
   blacklist.append(x)\;  
   x = DFT\_relaxation(x)\;
   blacklist.append(x)\;
   population.append(x)\;
   i+=1\;

 }
 \# iteration \;
 \While{j $<$ iterations}{
 	population.sort(index=energy)\;
	(parent1, parent2) = population.select\_candidates(2) \;
	(child1, child2)  = sensible\_crossover(parent1, parent2)  \;
	
  (child1, child2) = mutation(child1, child2) \;
  
 \Repeat{child1 and child2 are sensible and are not in the blacklist}{(child1, child2) = mutation(child1, child2)} \;
   blacklist.append(child1, child2)\;
   (child1, child2) = DFT\_relaxation(child1, child2)\;
   blacklist.append(child1, child2)\;
   population.append(child1, child2)\;
   population.sort(index=energy)\;
   population.delete\_high\_energy\_candidates(2)\;
	   \eIf{convergence criteria met}{
  break\;
   }{
   j+=1\;}
 }
\caption{Genetic algorithm for sampling the conformational space of molecules.}
\end{algorithm}

\subsubsection{Initialization of the population}

First, a random 3D structure is generated with RDKit directly from the SMILES code. This structure serves as a template for the upcoming geometries. Next, two lists of random values are generated: one for the rotatable bonds and one for the \textit{cis/trans} values. If the resulting 3D geometry is sensible, the structure is then subjected to local optimization. To generate an initial population of size N (\textbf{popsize}), N sensible geometries with randomly assigned values for torsion angles need to be built and locally optimized. The optimized geometries constitute the initial population. Due to the fact that the geometries are created one after another, all randomly built structures can but do not have to be made unique in order to increase the diversity of the initial population.
\subsubsection{An iteration of the GA}

Our GA follows the established generation-based approach, i.e. the population evolves over subsequent generations. After completion of the initialization, the first iteration can be performed. For this purpose, the population is sorted and ranked based on the total energy values $E_{i}$ of the different conformers  $i=1,...,N$. For each individual the fitness $F_{i}$ is being calculated according to:  

\begin{equation}
   F_{i} = \frac{E_{max}-E_{i}}{E_{max}-E_{min}}
      \label{eq:formel}
\end{equation}

$E_{max}$ is the highest energy and $E_{min}$ is the lowest energy among the energies of the conformers belonging to the current population. As a result, $F=1$ for the 'best' conformer and  $F=0$ for the 'worst' conformer. In the case of a population with low variance in energy values ($E_{max}-E_{min}<\textbf{energy\_var}$, default=0.001 eV ), all individuals are assigned a fitness of 1.0.

\textit{Selection}. Two individuals need to be selected prior to the genetic operations. We implemented three mechanisms for the selection.

(i) In the (energy-based) \textit{roulette wheel}\cite{Goldberg1989}, the probability $p_{i}$ for selection of a conformer $i$ is given by:

\begin{equation}
   p_{i} = \frac{F_{i}}{\sum_{n=1}^N F_{i}}
      \label{eq:formel2}
\end{equation}

With this, the probabilities of the conformers are mapped to segments of a line of length one. Next, two random numbers between zero and one are generated and the conformers whose segments contain these random numbers are selected. In the case when the sum of the fitness values is lower than a defined threshold near one (\textbf{fitness\_sum\_limit}, default=1.2) the best and a random individual are selected. 

(ii) The \textit{reverse roulette wheel} proceeds similarly to the \textit{roulette wheel} mechanism with the difference that the fitness values are swapped, i.e. new fitness $F^{*}_{i}$ is assigned to each conformer:
\begin{equation}
   F^{*}_{i} = F_{N+1-i}
      \label{eq:formel3}
\end{equation}

Analogously, the probability $p_{i}$ for selection of a conformer $i$ is given by:

\begin{equation}
   p_{i} =\frac{F^{*}_{i}}{\sum_{n=1}^N F^{*}_{i}}
      \label{eq:formel4}
\end{equation}

(iii) In the \textit{random selection} mechanism all individuals have the same chance to be selected. 

In all selection mechanisms the selected individuals must be different from each other so that the crossing-over has a chance to produce unique conformers. 

\textit{Crossing-over}. Crossing-over is considered to be the main feature distinguishing evolutionary algorithms from Monte Carlo techniques where only a single solution can evolve. Crossing-over allows the algorithm to take big steps in exploration of the search space \cite{Damsbo2004}. In our algorithm, a crossing-over step is performed if a generated random number (between zero and one) is lower than a defined threshold (\textbf{prob\_for\_crossing}, default=0.95). Between the selected individuals, parts of the representing vectors are then exchanged. To that end, the vectors characterizing the structure of both individuals are "cut" at the same single position (determined at random). The first part of the first individual is then combined with the second part of the second, and vice versa (a scheme explaining the crossing-over procedure is provided in Figure S1). Crossing-over is successful only when the resulting vectors can be used for generating sensible geometries. Otherwise the crossing-over is repeated until sensible geometries are generated or a maximum number of attempts (\textbf{cross\_trial}, default=20) is exceeded. In the latter case, exact copies of the selected conformers are used for the following step. 

\textit{Mutations} are performed independently for the values of \textit{cis/trans} bonds and of the rotatable bonds and if randomly generated numbers exceed corresponding thresholds (\textbf{prob\_for\_mut\_cistrans}, default=0.5; \textbf{prob\_for\_mut\_rot}, default=0.5). For each, the number of mutation events is determined by a randomly picked integer number not higher than the user-defined maximal number of allowed mutations (\textbf{max\_mutations\_cistrans} and \textbf{max\_mutations\_torsions}). For each mutation, a random position of the vector is determined and the mutation is chosen to affect the value of that variable. In case of \textit{cis/trans} bonds, the selected value is changed to $0^\circ$ if it was above $90^\circ$ or below $-90^\circ$, else it is changed to $180^\circ$. A selected rotatable bond is changed to a random integer between $-179^\circ$ and $180^\circ$. A mutation step is only successful if the geometry built after the mutation of the vector is sensible and unique. Otherwise the entire set of mutations in a mutation step is repeated until a sensible and unique structure is generated or a maximum number of attempts (\textbf{mut\_trial}, default=100) is exceeded. In this case, the algorithm terminates. The mutation is performed for both vectors generated via crossing-over.

\textit{Local optimization and update}. As the computational cost of the local optimization is significantly higher than all of the other operations \cite{Cheng2004, Tipton2013}, only unique and sensible structures are subject to local optimization. The structures are transferred to an external program for local geometry optimization (here: FHI-aims \citep{Blum2009, Havu2009}, see section DFT calculations). The application of local optimization was shown to facilitate the search for minima by reducing the space the GA has to search \cite{Wales1997, Johnston2003}. Thus, the implemented GA is closer to Lamarckian than to Darwinian evolution, as the individuals evolve and pass on acquired and not inherited characteristics. Afterwards, the population is extended by the newly optimized structures and, after ranking, the two individuals with highest energy are removed in order to keep the population size constant. 

\textit{Termination} of the algorithm is reached if one of the convergence criteria is met:
(i)~the lowest energy has not changed more than a defined threshold (\textbf{energy\_diff\_conv}, default=0.001 eV) during a defined number of iterations (\textbf{iter\_limit\_conv}, default=10), or (ii)~the lowest energy has reached a defined value (\textbf{energy\_wanted}), or (iii)~the maximal number of iterations (\textbf{max\_iter}, default=10 ) has been reached. The convergence criteria are checked only after a defined number of iterations (\textbf{iter\_limit\_conv}, default=10).

We are interested not only in finding the global minimum but also in finding low-energy local minima as many of them may be relevant. Thus, all of the generated conformers are saved and are available for final analysis even if only a subset of them constitutes the final population. Table~\ref{tbl_params} summarizes  practical GA parameters that were employed for one of the reference systems (isoleucine dipeptide). 

The parameters listed in Table~\ref{tbl_params} can be taken as indicative of settings that will work for many small to mid-size molecules. A few exceptions apply. Specifically, the \textbf{max\_iter} and the \textbf{popsize} parameters are set to low values in Table~\ref{tbl_params}, covering only a small set of structures within an individual GA run. This choice would be appropriate for an ensemble of many short independent GA run to generate a broad structural ensemble with a bias towards the low-energy solution space. For larger and more complex molecules, and/or for runs designed to identify the global minimum in a single shot, \textbf{max\_ite}r could be increased significantly, and \textbf{popsize} could be increased somewhat (to 10-20 individuals) as well. Likewise, the mutation probabilities \textbf{prob\_for\_mut\_cistrans} and \textbf{prob\_for\_mut\_rot} are here set to relatively high values of 0.5, instilling a significantly amount of randomness into the search process. For a more "deterministic" search process, somewhat smaller values (e.g., 0.2) might be chosen. Finally, the \textbf{distance\_cutoff} criteria are chosen to be appropriate for light elements (first and second row); adjustments may be appropriate if heavier covalently bonded atoms are included in the search.

\begin{table}
  \caption{GA parameters for isoleucine dipeptide}
  \label{tbl_params}
   \small
\begin{tabular}{|p{2.9cm}|p{5.0cm}|p{7.5cm}|}
    \hline
    & Parameter &  Value \\
     \hline
     \hline
     \multirow{5}{*}{Molecule} & SMILES  & \footnotesize{CC(=O)N[C@H](C(=O)NC)[C@H](CC)C} \\
    & distance\_cutoff\_1 & 1.2  \r{A}  \\
    & distance\_cutoff\_2 & 2.0  \r{A}  \\
    & rmsd\_cutoff\_uniq & 0.2  \r{A}  \\
	& chiral & True \\       
    \hline    
    \multirow{3}{*}{Run settings} & max\_iter  & 10  \\
	& iter\_limit\_conv & 10  \\
    & energy\_diff\_conv & 0.001 eV\\ 
        \hline     
    \multirow{11}{*}{GA settings} &  popsize & 5 \\
    &energy\_var & 0.001 eV\\
    & selection & roulette wheel \\
	& fitness\_sum\_limit & 1.2\\   
    & prob\_for\_crossing & 0.95  \\
    & cross\_trial & 20 \\
    & prob\_for\_mut\_cistrans & 0.5 \\
    & prob\_for\_mut\_rot & 0.5 \\
    & max\_mutations\_cistrans & 1    \\
    & max\_mutations\_torsions & 2   \\
    & mut\_trial & 100 \\
   
    \hline
  \end{tabular}
\end{table}

\subsubsection{DFT calculations}

For the tests presented below, all DFT calculations are performed with the FHI-aims code \cite{Blum2009, Havu2009}. We employed the PBE functional \cite{Perdew1996} with a  correction for van der Waals interactions (pairwise\cite{Tkatchenko2009}  for the amino acid dipeptides calculations and MBD \cite{Ambrosetti2014} for the drug-like ligands) and with \textit{light} computational settings and \textit{tier} 1 basis set\cite{Blum2009}. For the local optimization, we use a trust radius enhanced version of the Broyden–Fletcher–Goldfarb–Shanno (BFGS) algorithm \cite{bfgs} initialized by the model Hessian matrix of Lindh \cite{Lindh1995}. This is the default choice in the FHI-aims code and was implemented by J\"{u}rgen Wieferink. The local optimization is set to terminate when the maximum residual force component per atom is equal to $5\cdot{10^{-3}}$  eV/ \r{A}. Density functional, basis set, and numerical settings (e.g. integration grids) are user choices of the underlying density-functional theory code and must be set appropriately outside of Fafoom. The settings for numerical convergence (including basis set) must be chosen converged enough but not introduce artifacts in the landscape of minima found. The choice of the density-functional approximation (DFA) to the exact Born-Oppenheimer potential-energy surface needs to reproduce the local energy minima of the PES faithfully, as discussed in the introduction. We here only note that costs for different electronic structure approximation can vary by orders of magnitude. In practice, and strictly speaking, the scope of our algorithm is to find the PES minima for a given DFT functional, while the physical choice of the "right" DFA is not the focus of this paper. We do show, however, that we can use our algorithm in practice with one specific functional, PBE functional with a correction for van der Waals interactions, that has yielded very reliable results in the past.

\subsubsection{Parallelization}
Parallel computational resources can be utilized in two ways in order to speed up the computation. First, multiple GA runs can be started in parallel and the blacklist can be shared between different and subsequent runs. Sharing the blacklist increases diversity of solutions with already a few GA runs. Second, the time needed for the individual energy evaluations can be decreased if the molecular simulation package allows calculations across distributed nodes and is efficiently parallelized (e.g. in FHI-aims \cite{Marek2014}). Our code supports both modes of computation.

\subsubsection{Availability of the code}
The code is distributed as a python package named Fafoom (Flexible algorithm for optimization of molecules) under the GNU Lesser General Public License\cite{gnuglgp}. It is available from following websites:
\begin{itemize}
\item https://aimsclub.fhi-berlin.mpg.de/aims\_tools.php
\item https://github.com/adrianasupady/fafoom
\end{itemize}
Although designed for usage with a first-principles method (e.g. FHI-aims, NWChem \cite{Valiev2010}), Fafoom can also be used with a force field (MMFF94\cite{Halgren1996}, accessed within RDKit\cite{rdkit, Tosco2014}) for testing purposes. It is in principle possible to use any molecular simulation package which outputs optimized geometries together  with their energies. Nevertheless, this requires adjusting a part of the program to the specific needs of the used software. Details are provided with the program’s documentation.  

\subsection{Reference data}

In order to evaluate the several aspects of the performance of the implemented algorithm we use two sets of reference data. The first reference data set (\textbf{Amino acid dipeptides}) was extracted from a database of computational data for the amino acid dipeptides. The second reference data set (\textbf{Mycophenolic acid}) contains conformers of a drug-like ligand that were obtained with three different search techniques.

\subsubsection{Amino acid dipeptides}

\begin{figure}
\includegraphics{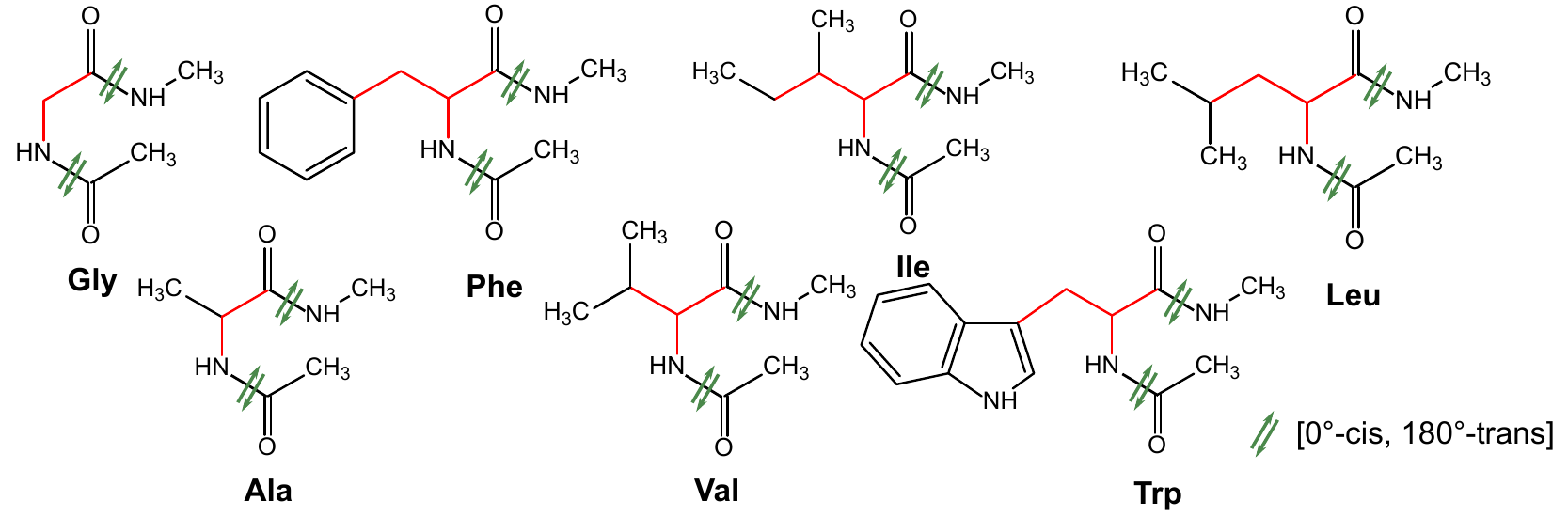}
 \caption{Chemical structures of the amino acid dipeptides. Rotatable bond are single, non-ring bonds between non-terminal atoms that are not attached to methyl groups that carry three identical substituents and are marked in red. Double arrows mark the \textit{cis/trans} bond.}
  \label{fig_chemstr}
\end{figure}

The first reference data set contains conformers of seven amino acid dipeptides \cite{dipeptide} (Figure~\ref{fig_chemstr}) and was extracted from a large database for amino acid dipeptide structures generated in a combined basin-hopping/multi-tempering based search. In that search (published in detail in \cite{Ropo2015}), the framework of the reference search can be divided into a global search step and a refinement step. In the global search, the basin hopping search technique together with an empirical force field OPLS-AA was employed to perform the initial scan of the PES. The identified energy minima were relaxed at the PBE+vdW level with \textit{light} computational settings in FHI-aims. In the refinement step, \textit{ab initio} replica-exchange molecular dynamics runs were performed to locally explore the conformational space and to alleviate a potential bias of the initial search of a force field PES. The resulting minima were again optimized at the PBE+vdW level with \textit{tight} computational settings and with the \textit{tier} 2 basis set\cite{Blum2009}. In order to compare to our data, they were re-optimized with the same functional with \textit{light} computational settings, and the \textit{tier} 1 basis set\cite{Blum2009}. After this procedure, duplicates were removed from the set used for the comparison with the GA results. For benchmarking the performance of our search strategy for conformers predictions, we consider all structures with a relative energy up to 0.4 eV. These conformers define the reference energy hierarchy for each of the selected dipeptides. We summarize some characteristics and the number of conformers that were considered in Table~\ref{tbl_amino}. 

\begin{table}
  \caption{Reference data set: seven amino acid dipeptides.}
  \label{tbl_amino}
   \small
\begin{tabular}{|p{2.9cm}|p{2.0cm}|p{2.0cm}|p{3.75cm}|p{3.75cm}|}
    \hline
     Amino acid dipeptide & Abbr. & No. of atoms  & No. of rotatable bonds + No. of \textit{cis/trans} bonds & No. of conformers (below 0.4 eV $\approx$ 38.6 kJ/mol) \\
     \hline
     \hline
   Glycine & Gly& 19 & 2+2  & 15 (15)  \\    
   Alanine & Ala& 22 & 2+2  & 28 (17)  \\   
   Phenylalanine &Phe& 32 & 4+2  & 64 (37)  \\
   Valine &Val& 28 & 3+2  & 60 (40)  \\
   Tryptophan &Trp& 36 & 4+2  & 141 (77)  \\
   Leucine &Leu& 31 & 4+2  & 183 (103)  \\
   Isoleucine & Ile& 31 & 4+2  & 176 (107)  \\
     
        \hline     
  \end{tabular}
\end{table}

\subsubsection{Mycophenolic acid}

\begin{figure}
\includegraphics{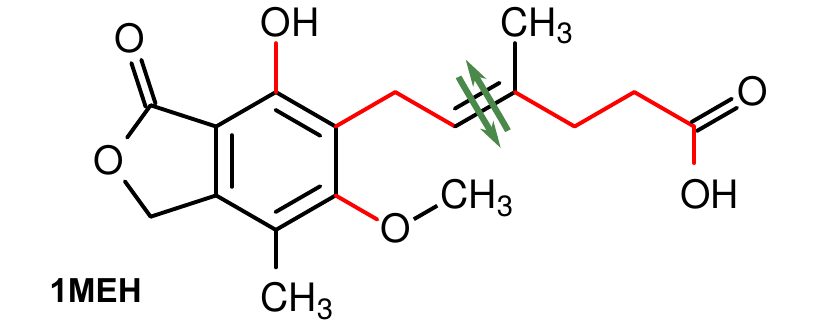}
 \caption{The chemical structure of the selected ligand together with the PDB-ID of the respective X-ray structure of the target protein. Rotatable bonds are marked in red and the \textit{cis/trans} bond is marked with double arrows.}
  \label{fig_str_1meh}
\end{figure}

From the Astex Diverse Set \cite{Hartshorn2007}, a collection of X-ray crystal structures of complexes containing ligands from the Protein Data Bank (PDB), one example for a drug-like ligand was selected. This molecule, mycophenolic acid (target protein: 1MEH) has 43 atoms, 8 rotatable bonds and 1 \textit{cis/trans} bond (Figure~\ref{fig_str_1meh}).

Mycophenolic acid is a very flexible molecule. Even a coarse systematic search with a grid of only $60^\circ$ for the freely rotatable torsions and 2 values (cis/trans) for the double bond and the X-X-O-H torsions yields already $6^{6}*2*2*2=373248$ conformations to test. This makes  this molecule a challenging example to test the performance of three search techniques (A-C below) in combination with first-principles methods.

A) \textit{Genetic algorithm}. 50 independent GA runs with following settings: \textbf{max\_iter}=30, \textbf{iter\_limit\_conv}=20 and  \textbf{popsize}=10, were performed with Fafoom. A total of 3208 structures were generated.

B) \textit{Random search}. 3200 random and clash-free structures were generated with Fafoom and further relaxed with DFT.

C) \textit{Systematic search with Confab \cite{O'Boyle2011b}}. First, 293 conformers were generated with Confab (assessed via Open Babel, used settings:  \textbf{RMSD cutoff} = 0.65 and \textbf{Energy cutoff} = 15 kcal/mol ). In order to account for two different values for the \textit{cis/trans} bond and the X-X-O-H torsions ($0^\circ$ and $180^\circ$), eight starting structures per each of the conformers generated with Confab were considered. This procedure yields overall 2344 structures. After removing geometries with clashes, \textbf{2094} structures were subjected to DFT optimization.

Finally, all DFT optimized structures were merged to a common pool and the duplicates were removed. For this, a two-step criterion was used. First, the compared structures need to have a torsional RMSD (tRMSD) lower than $0.1\pi$ rad \cite{trmsd}. Second, the energy difference between the compared structures cannot exceed 10 meV. If both criteria are met, the structure that is higher in energy is labelled as 'duplicate' and is removed from the pool.
In total, 1436 unique structures were found. Table S1 shows the number of the obtained unique structures depending on the applied energy cutoff. 

\section{Results and discussion}

The performance of the GA search is evaluated by the ability to reproduce the reference energy hierarchies and to find the global minimum. We performed multiple GA runs for the test systems to test the impact of varying search settings.

\subsection{Amino acid dipeptides}

For each of the amino acid dipeptides we performed 50 independent GA runs with 10 iterations (\textbf{max\_iter}) each and a population size of 5 (\textbf{popsize}). One GA run with such settings requires $\textbf{popsize}+2\cdot\textbf{iterations}=25$ geometry optimizations at the PBE+vdW level and yields 25 conformers.

\subsubsection{Finding the global minimum}

First we assess the probability to find the global minimum (known from the reference energy hierarchy) among them. We check how many of the GA runs succeed in finding the global minimum and subsequently calculate the probability for finding the global minimum in one GA run and present the results in Table~\ref{glob_min_both}.

\begin{table}[h!]
  \caption{Average (from 50 GA runs) probability for finding the energy global minimum in a given run with 25 locally optimized conformers.}
  \label{glob_min_both}
   \small
\begin{tabular}{|p{4.0cm}|p{1.3cm}|p{1.3cm}|p{1.3cm}|p{1.3cm}|p{1.3cm}|p{1.3cm}|p{1.3cm}|}
  
    \hline
     Molecule 	& Gly & Ala & Phe & Val & Trp & Leu & Ile \ \\
     \hline
         \hline
     TDOFs 		& 4 & 4 & 6 & 5 & 6 & 6 & 6  \\
Probability for global minimum (/1 run) & 0.82 & 0.79 & 0.53 & 0.60 & 0.22 & 0.20 & 0.10 \\
       \hline
  \end{tabular}
\end{table}

Table~\ref{glob_min_both} illustrates how the magnitude of the sampling problem does not only depend on the dimensionality, i.e. here the number of TDOFs, but also on the chemical structure.
Phenylalanine and isoleucine are two interesting cases, both have the same number of TDOFs and are of similar size, but the probability of finding the global minimum with a single run drops dramatically. The drop in probability is, of course, correlated with the overall number of conformers listed in Table~\ref{tbl_amino}. 

\subsubsection {Conformational coverage}

\begin{figure}[h!]
\includegraphics{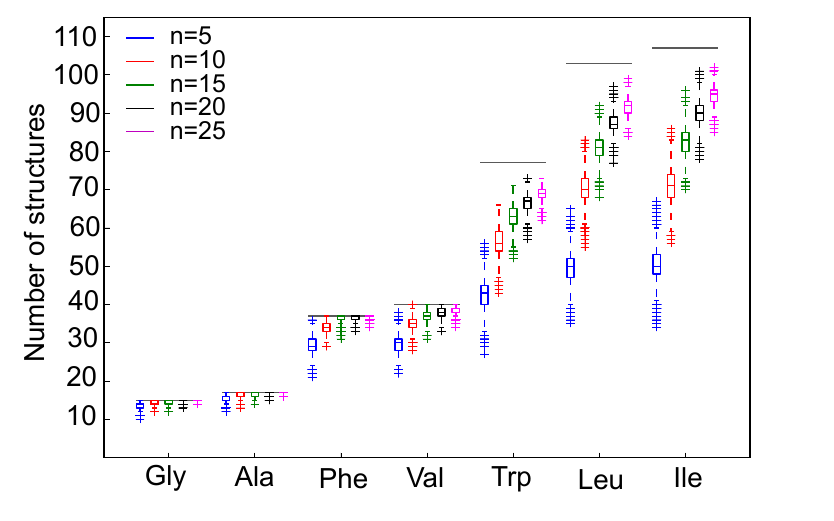}
 
  \caption{Number of minima found by the GA for seven amino acid dipeptides. The horizontal lines depict the total number of minima for the given compound as predicted by Ropo \textit{et al.} \cite{Ropo2015}.
From a total of 50 GA runs, 5, 10, 15, 20, 25 GA runs were randomly selected and the found structures were counted. This procedure was repeated 10,000 times and the resulting distributions are summarized in box plots. The line inside the box is the median, the bottom and the top of the box are given by the lower ($Q_{0.25}$) and upper ($Q_{0.75}$) quartile. The length of the whisker is given by $1.5\cdot{(Q_{0.75}-Q_{0.25})}$. Outliers (any data not included between the whiskers) are plotted as a cross.}
  \label{hierar_all}
\end{figure}

A key point in our approach is to reproduce the known energy hierarchies of the reference systems. For each of the investigated compounds, we randomly choose 5, 10, 15, 20, and 25 runs (from the pool of 50 runs), merge the results, and check how many structures have been found. We repeat this procedure 10,000 times and present the results in Figure~\ref{hierar_all}.

\begin{figure}[h!]
\includegraphics{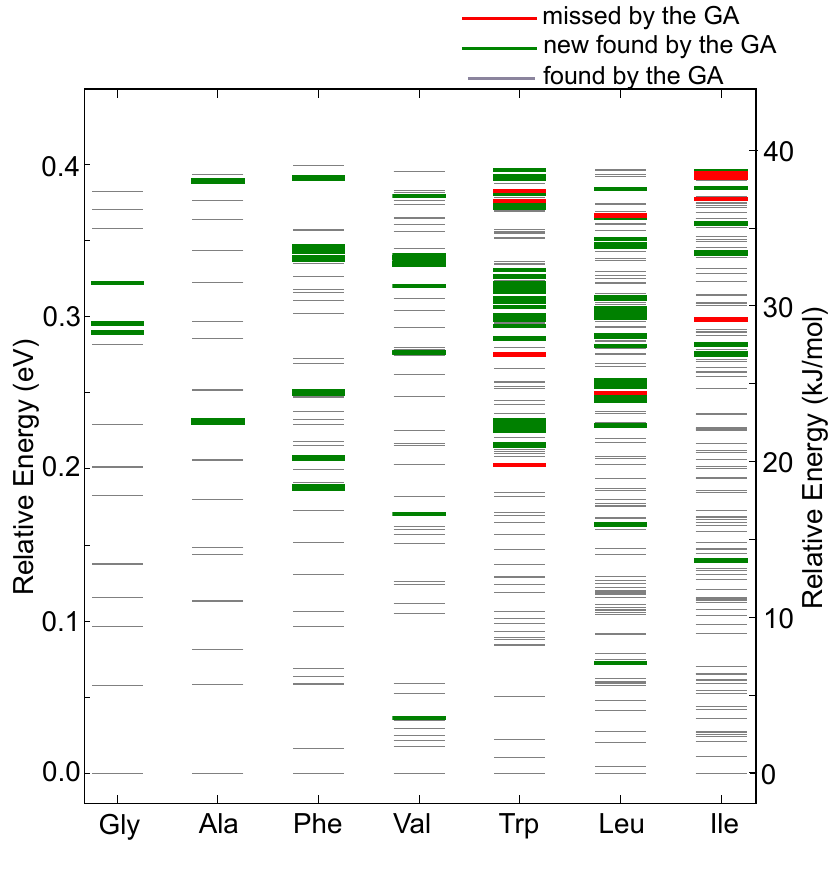}
 \caption{Difference hierarchies for the amino acid dipeptides. Red lines depict structures from the reference data set that have not been found by the GA. Green lines depict structures found by the GA that were absent in the reference data set. Gray lines depict structures from the reference data set that were found by the GA. The results from all 50 GA runs for each dipeptide were taken into account.
 }
  \label{diff_hierarchy}
\end{figure}

It is evident that for dipeptides with a small number of reference minima (alanine and glycine) we obtain a very good results, i.e. a very good coverage of conformational space, already with 5 repeats of the GA runs. For dipeptides with a slightly higher number of minima (phenylalanine and valine) at least 10 runs of the GA are needed to obtain a good result. For the remaining dipeptides, the GA is not able to find all of the reference minima, even with 25 GA runs. However, the coverage of the reference hierarchy with 20 GA runs is always higher than 80\%. We next inspect in more detail which of the amino acid dipeptides' reference minima were missed. To this end we investigate the actual difference between the reference hierarchy and the hierarchy obtained from the 50 GA runs, see Figure~\ref{diff_hierarchy}.
Although our search strategy misses a few of the reference structures even when 50 repeats of the GA search are performed, the first missed structure has a relative energy higher than 0.2 eV. This in turn means that no low-energy structures are being missed. Furthermore, there are multiple newly predicted structures that were not present in the reference data set (Figure~\ref{diff_hierarchy}). It should be noted that, considering the fact that the investigated GA runs are rather short, the random component of the search (randomly initialized populations) contributes to the good results of the search.

\subsubsection{Parameter sensitivity}

In order to check the robustness of the default run parameters, several alternative settings were tested for the isoleucine  dipeptide. The tested parameters include: (i)~the impact of the selection mechanism (roulette wheel, reverse roulette wheel, random), (ii)~the effect of decreasing the cut-off for blacklisting from the default value of 0.2\,\r{A} to 0.05\,\r{A}, and (iii) the increase of the maximal number of iterations from the default 10 to 15, 20 and 25. For cases (i) and (ii) 100 GA runs were performed for each of the settings. In order to assess the effect of the number of iterations, 100 runs with a maximal number of iterations equal to 25 have been performed and subsequently only considered up to a maximum of 15, 20, 25 maximum iterations. Additionally, 50 GA runs with a maximal number of iterations equal to 100 were performed. In all mentioned cases convergence criteria were evaluated after each iteration, starting from the \textbf{iter\_limit\_conv}=10th iteration. 

We find that none of the three selection mechanisms has a distinct impact on the probability for finding the global minimum or quality of the conformational coverage. Similarly, no substantial change was observed upon the decrease of the blacklisting cut-off. The probability value for finding the global minimum as well as the number of found reference minima increases with increased number of iterations. This is simply due to the increased number of trials for sampling the conformational space. Table~\ref{glob_min_sensitivity} summarizes the probability to find the global minimum in one run with different settings. Detailed data about the conformational coverage is given in Figure S2. 
 
\begin{table}[h!]
  \caption{Probability of finding the global minimum of isoleucine in one run for different setups. The default settings include roulette wheel selection mechanism, 0.2 \r{A} cut-off for the blacklisting and maximal number of iteration equal to 10. The numbers in brackets denote the mean number of iterations needed for convergence.}
  \label{glob_min_sensitivity}
  
\small  
  \begin{tabular}{|p{5.3cm}|p{5.3cm}|p{4.6cm}|}
    \hline

\multicolumn{2}{|c|}{Setup} & Probability of finding the global minimum (per run) \\
\hline 
\hline
\multicolumn{2}{|c|}{default} & 0.17 \\
\hline
\multirow{2}{*}{Selection mechanism} & roulette wheel reverse & 0.18 \\
& random & 0.13 \\
\hline 
\multirow{4}{*}{Max. number of iterations} & 15 (13) & 0.20 \\
& 20 (15) & 0.25 \\
& 25 (16) & 0.25 \\
& 100 (22) & 0.46 \\
\hline
Cut-off for blacklisting & 0.05 \r{A} & 0.14 \\
\hline

  \end{tabular}
\end{table}

\subsubsection{Evaluation of the computational performance}

The accuracy of a search/sampling strategy is its most crucial feature. Nevertheless, its computational cost plays a significant role in practical applications. To this end, we quantify the total cost of the GA runs in terms of force evaluations required in the local geometry optimizations. The number of force evaluations, i.e. most expensive steps in the algorithm,  is a suitable measure for the computational cost. One force evaluation requires approximately between 1 (glycine) to 3 (tryptophan) CPUminutes. We quantify the number of force evaluations required by the GA for reproducing 85\% of the reference hierarchy and present the results in Table~\ref{tbl_time_comp_aa}.  The table also includes the number of force evaluations required only in the replica-exchange MD refinement step of the reference search (the number of force evaluations required for the geometry optimizations is not even included).

\begin{table}[h!]
  \caption{Comparison of the computational cost: amino acid dipeptides. The cost is given in the total number of force evaluations [$*10^{3}$].}
  \label{tbl_time_comp_aa}
\noindent 
\small
\begin{tabular}{|p{5.05cm}|p{1.15cm}|p{1.15cm}|p{1.15cm}|p{1.15cm}|p{1.15cm}|p{1.15cm}|p{1.15cm}|}
  
    \hline
      & \multicolumn{7}{|c|}{Total number of force evaluations[$*10^{3}$]}\\

      	& Gly & Ala & Phe & Val & Trp & Leu & Ile \ \\
     \hline
         \hline
     
GA (at least 85\% reproduction of the reference hierarchy) & 11 & 12 & 29 & 24 & 60 & 68 & 61 \\
Reference & 380 & 400 & 480 & 440 & 500 & 460 & 460 \\
       \hline
  
  \end{tabular}
\end{table}

\subsection{Mycophenolic acid}

\begin{figure}
\includegraphics{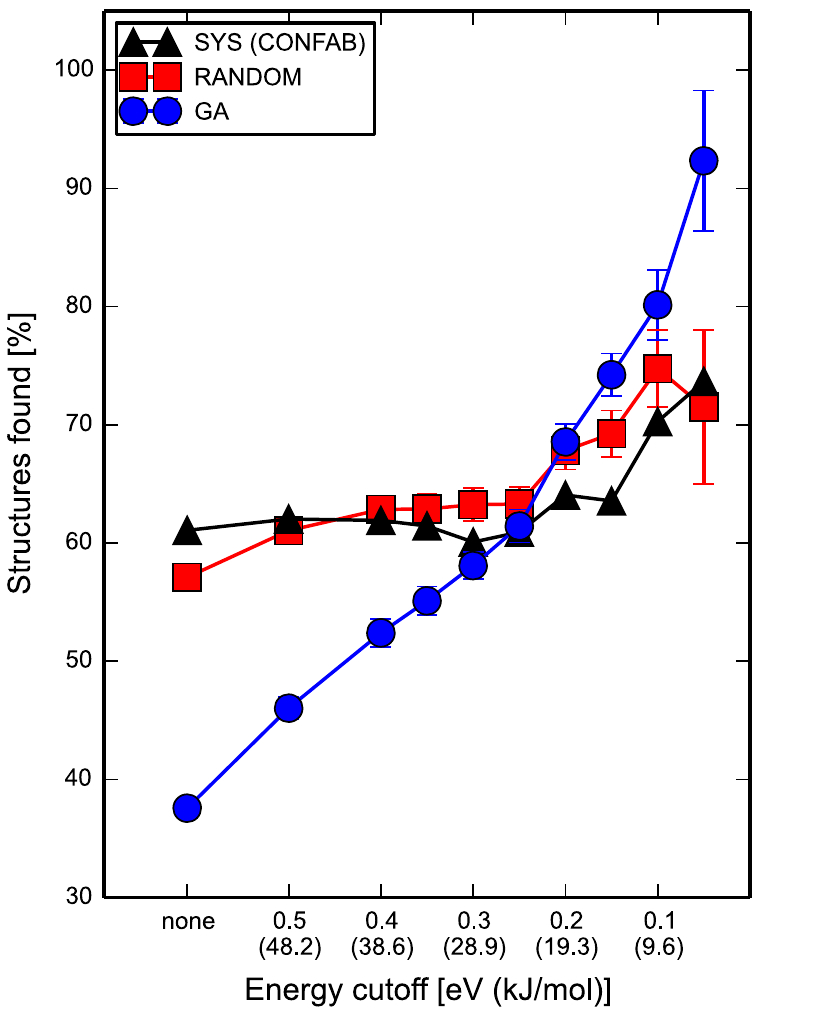}
 
  \caption{Share of the reference number of structures found by three search techniques: GA (blue circles), random search (red squares) and systematic search with Confab (black triangles) as a function of the applied energy cutoff. Energy values are given in eV and in parentheses also in kJ/mol.}

  \label{1meh_compare_fig}
\end{figure}

In the following we utilize as reference a set of structures that is a result of merging all structures found by three techniques: 3208 structures from 50 GA runs, 3200 random structures and 2094 structures generated with Confab. We define the following subsets: (i) 'GA' is a random selection of 25 GA runs (approx. 1600 structures); (ii) 'SYS (CONFAB)' is a set of all 2094 structures generated in the systematic search; and (iii) 'RANDOM' is random selection of 1600 structures generated in the random search. For the performance evaluation we count how many of the reference structures can be found by the respective search technique. This procedure was repeated 1000 times for each of the energy cutoffs. The results are shown in Figure \ref{1meh_compare_fig}. More details can be found in Table S1.

All of the search techniques found the same global minimum several times. In case if no energy cutoff is applied, none of the searches is able to find all local minima in the conformational space (i.e. more calculation would be needed). With a decreasing energy cutoff, an improved coverage of the conformational space can be observed. The fact that the GA is a global optimization techniques is clearly visible as it performs better in the low-energy ($<0.2$ eV) region, whereas the random and systematic search perform uniformly but not perfectly independent of the energy cutoff used for the evaluation.

In order to show the wide and routine applicability of our first-principles structure search approach, we have performed short exploratory structure searches (only 3 GA runs each) to eight drug-like ligands from the Astex Diverse Set, which is widely utilized to assess the performance of, for example, conformer generators. The molecules vary in the size (8 to 32 atoms) and number of rotatable bonds (6 to 13). A detailed analysis of this study is shown in the Supporting Information. In brief we find that in all eight instances a diverse pool of conformers can be generated. In each case, a conformer is found that is similar to the protein-bound ligand from the X-ray structure with an RMSD of 1.5 \r{A}. In six of the eight instances, they are similar with RMSD values of less than 0.9 \r{A}. Exploratory first-principles structure searches have a potential application in \textit{in silico} protein-ligand studies\cite{Avgy-David2015}: the comparison of the structural space of the isolated ligand and the structure realized by the protein-bound ligand might reveal details about the binding process, for example whether the binding mechanism follows more the conformational-selection or induced-fit type. In contrast to many of the quicker (but simpler) established conformer generators, the first principles energetics that we obtain here are not dependent on initial parametrizations and thus the method is in principle applicable throughout chemical space. It is important to note that, in this test, our goal was not to provide a converged GA search for each molecule but rather to explore the GA's potential to provide approximate conformational coverage with a fixed computational budget. Our investigation of mycophenolic acid indicates that searches for each of these molecule could be reliably converged albeit at significantly higher computational expense.

\subsubsection{Literature context}
In order to put the algorithm's parameters into perspective, we compare it to four selected  applications of EA or GA to the conformational search of molecules in the following. In all considered algorithms,  the initial populations are generated randomly and the conformational space of the respective molecules is represented and sampled (by mutation and crossing-over) by means of torsion angles, i.e. rotations around bonds. Table~\ref{alg_comp} summarizes a few parameters that illustrate the range over which the parameters that are characteristic to these kinds of evolutionary or genetic algorithms can vary. The approaches differ in the energy functions that are employed: Damsbo \textit{et al.} \cite{Damsbo2004} employ the CHARMM force field,  Vainio and Johnson \cite{Vainio2007} use the torsional and the vdW term of the MMFF94 force field separately in a multi-objective genetic algorithm (MO-GA) while Nair and Goodman\cite{Nair1998} use the MM2* force field. The study on optimizing the GA parameters for molecular search with a meta-GA, presented by Brain and Addicoat\cite{Brain2011}, uses, similar to our work, a first-principles energy functions. Two choices in the algorithm highlight the difference between theirs and our aim: in order ''to reliably find the already known \textit{a priori} correct answer with minimum computational resources``, the selection criterion 'rank' focuses on the generation's best solution. Furthermore, crossing-over is considered as not helpful. In contrast, the aim of our work is to provide a GA implementation that ensures broad conformational coverage, i.e. the prediction of an energy hierarchy and not only the reproduction of a global optimum. For that we found it useful to employ random or roulette-wheel selection that also accepts less-optimal structures for genetic operations and a high probability for crossing-over. Both choices (accompanied by blacklisting) can be interpreted as means to increase diversity during the search.

\begin{table}[h!]
  \caption{Comparison of parameters and schemes that are used in search approaches proposed in four selected publications with the approach presented in this work.}
  \label{alg_comp}
  
\small    
\begin{tabular}{|p{4.2cm}|p{1.95cm}|p{1.95cm}|p{1.95cm}|p{1.95cm}|p{1.95cm}|}

\hline
 Parameter &  Damsbo'04 \cite{Damsbo2004} & Vainio'07 \cite{Vainio2007}  & Nair'98 \cite{Nair1998} & Brain'11 \cite{Brain2011}& this work \
\\

\hline
\hline
Algorithm type & EA & MO-GA & GA & GA & GA\\
Population size & 30 & 20 & 2-20 & 10-15 & 5 \\
Selection & - & tournament & roulette & rank & roulette \\
Crossing-over probability & - & 0.9 & 1.0 & 0.0-1.0 & 0.95  \\
Mutation probability & - & 0.05 & 0.4 & 0.3-0.5 & 0.5\\

\hline
\end{tabular}
\end{table}

\section{Conclusions}

We aimed at designing a user-friendly framework with an implementation of the genetic algorithm for searches in molecular conformational space that is particularly suitable for flexible organic compounds. A SMILES code for the selected molecule is the only required input for the algorithm. Furthermore, a wide selection of parameters (e.g. torsion definition, blacklist cut-off)  allows for customizing the search. With minor changes, the code can be interfaced to external packages for molecular simulations that output optimized geometries together with corresponding energies. Besides its adaptability and ease of use, a further advantage of the implementation is the fact that it allows for using first-principles methods. With this, a potential bias resulting from the parametrization of a particular force-field can be avoided and makes the search applicable to a broad selection of problems. We examined the performance of the implementation in terms of efficiency and accuracy of the sampling. The algorithm is capable of reproducing the reference data with a high accuracy. For a set of amino acid dipeptides, we show that this conformational coverage is achieved much more efficiently than in an earlier, \textit{ab initio} replica-exchange MD based search in our group. For a larger molecule (mycophenolic acid), we show that the low-energy conformational space coverage of the GA surpasses the coverage of two competing methods significantly at similar effort.

\acknowledgement{Matthias Scheffler (FHI Berlin) is kindly acknowledged for support of this work and scientific discussions.}

\section{Supporting Information}

\subsection{Crossover procedure}

\begin{figure}
\includegraphics{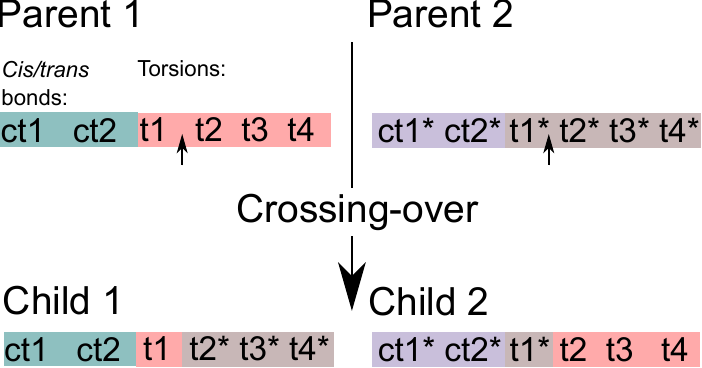}
 
  \caption{Crossing-over procedure. The lists of values for \textit{cis/trans} bonds and torsions are first combined together. As next, a single cut is performed and the corresponding parts are exchanged.}
\label{cross_over}
\end{figure}

The following scheme (Figure~\ref{cross_over}) explains the crossover procedure described in the \textbf{Methods} section of the article.

\subsection{Conformational coverage: Isoleucine}

\begin{figure}

\includegraphics{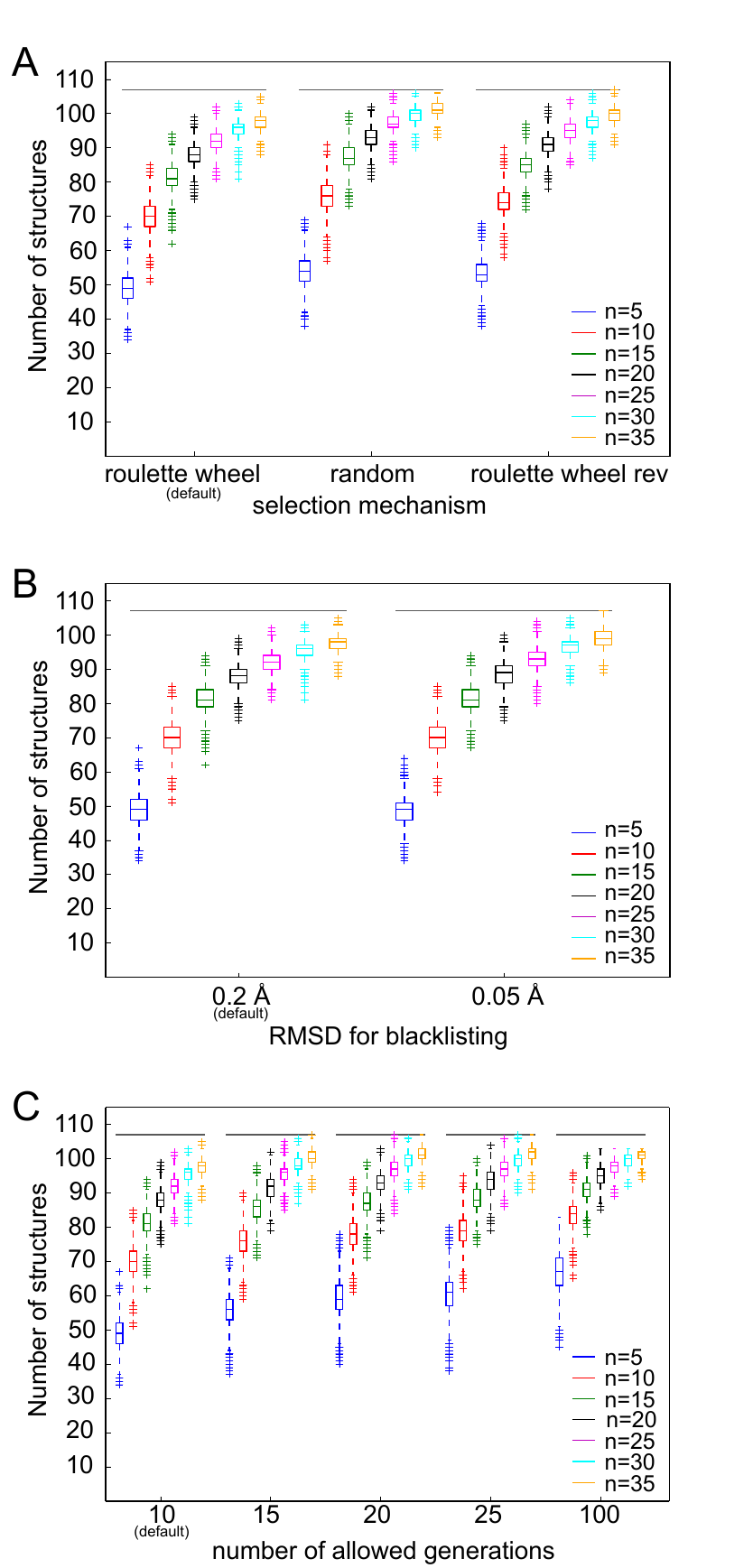}
 
  \caption{Conformational coverage of the GA search with different settings for Ile. The horizontal lines depict the number of reference minima\cite{Ropo2015} (107). From a total of 100 GA runs, 5, 10, 15, 20, 25, 30, 35 GA runs were randomly selected and the found structures were counted. This procedure was repeated 10,000 times and the resulting distributions are summarized in box plots. The line inside the box is the median, the bottom and the top of the box are given by the lower ($Q_{0.25}$) and upper ($Q_{0.75}$) quartile. The length of the whisker is given by $1.5\cdot{(Q_{0.75}-Q_{0.25})}$. Any data not included between the whiskers is plotted as an outlier with a cross. Conformational coverage hardly changes by using different selection mechanisms (A) or changing the blacklisting cut-off (B). (C)~Increasing the number of GA iterations improves conformational coverage.}
  \label{sensitivity_box}
\end{figure}

Figure~\ref{sensitivity_box} shows the coverage of the conformational space of isoleucine dipeptide resulting from GA searches with different settings.

\subsection{Conformational coverage: Mycophenolic acid}

Table~\ref{count_three} contains the total number of mycophenolic acid conformers found by three conformer generation techniques depending on the energy cutoff. Further, it contains detailed data used to create Figure 6 in the article. 
As there are three search techniques there are eight possible cases for finding a structure or not: (i) a structures can be found by all searches ('all') or (ii) a structure can be found by two of three searches ('\textit{GA+RANDOM}','\textit{GA+SYS}', '\textit{SYS+RANDOM}') or (iii) a structure can be found by only one search ('\textit{only GA}', '\textit{only RANDOM}', '\textit{only SYS}') or (iv) a structure can be missed by all of the searches (not included in the table).

\begin{table}[h!]
  \caption{Share of found structures by the GA, systematic search and random conformer generation depending on the used energy cutoff.}
  \label{count_three}
   \small
\begin{tabular}{|p{1.76cm}|p{1.05cm}|p{1.05cm}|p{1.1cm}|p{1.3cm}|p{1.3cm}|p{1.3cm}|p{1.3cm}|p{1.3cm}|p{1.3cm}|p{1.1cm}|}
    \hline
     Energy cutoff (eV) / 
     (kJ/mol) &  Nr. of structures & Unique structures &\multicolumn{7}{|c|} {[\%] of structures found by } \\
     & & & only GA&only RANDOM&only SYS & GA + RANDOM & GA + SYS & RAN- DOM + SYS & all  \\
     \hline
     \hline
no & 8502 & 1436 & 6.28 &14.00 &19.23 &6.45 &5.20 &17.00 &19.65\\
0.5 (48.2)& 7164 & 1006 &7.18& 12.04& 15.26& 8.13& 5.90& 16.05& 24.82\\
0.4 (38.6) & 6288 & 764 & 8.02 &10.74 &12.95 &9.30 &6.17 &13.89 &28.90  \\
0.35 (33.8)& 5452 & 625 &8.43& 10.30 &12.69 &10.22 &6.39 &12.29 &30.07 \\
0.3 (28.9)& 4961 & 531 &9.10 &9.97 &11.53 &11.36 &6.62 &10.92& 31.00  \\
0.25 (24.1) & 4535 & 458 &9.53 &8.39& 11.23& 12.38& 7.16 &10.16 &32.37   \\
0.2 (19.3)& 4079 & 345 & 8.53 &6.79& 9.52& 14.27 &7.78 &8.79 &37.96 \\
0.15 (14.5) &3153& 225 & 9.51 &5.64 &8.17& 15.71& 7.49& 6.37& 41.53 \\
0.1 (9.6) & 1314 & 74& 7.04& 4.33& 6.56& 14.07& 7.36 &4.67 &51.68\\ 
0.05 (4.8)& 298 & 19 &6.87& 2.81& 1.04& 15.10& 19.05& 2.28& 51.31\\

    \hline
  \end{tabular}
\end{table}

\subsection{A small set of flexible organic molecules}

We utilize the Astex Diverse Set \cite{Hartshorn2007}, a collection of structures obtained from X-ray crystal structures from the Protein Data Bank (PDB), to construct the a small set of flexible organic molecules.  
The goal here is not to provide a converged GA search for each molecule but rather to explore the GA's potential to provide approximate conformational coverage with a fixed computational budget. Our investigation of mycophenolic acid indicates that searches for each of these molecule could be reliably converged albeit at significantly higher computational expense.
 
\begin{figure}
\includegraphics{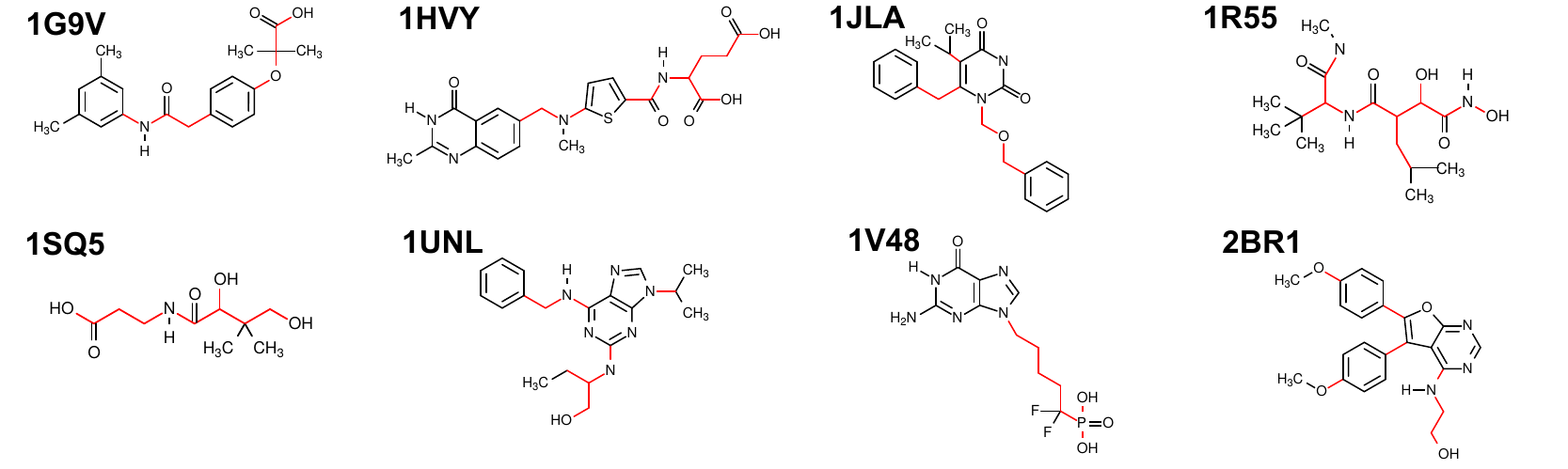}
 \caption{Chemical structures of the selected ligands together with the PDB-IDs of the respective X-ray structures of the target proteins. Rotatable bonds are marked in red.}
  \label{fig_str_astex}
\end{figure}

We selected 8 ligands (Figure~\ref{fig_str_astex}) that differ in composition, the number of heavy atoms (15-32) and the number of rotatable bonds (6-13)\cite{torsionOH}. 

\subsubsection{Genetic algorithm searches}
\textbf{SMILES} codes for the respective entries were taken from PubChem to ensure an unbiased starting point for the search. We utilize the parameter values as listed in Table 2 in the article with following exceptions: \textbf{max\_iter}=30, \textbf{iter\_limit\_conv}=20, \textbf{popsize}=10, \textbf{prob\_for\_mut\_rot}=0.8, \textbf{prob\_for\_mut\_cistrans}=1, \textbf{cross\_trial}=100 and \textbf{max\_mutations\_torsion}=3. For each of the molecules, three GA runs have been performed, i.e. given the settings, the number of obtained conformers cannot be higher than 210.
\subsubsection{Results}
\begin{figure}
\includegraphics{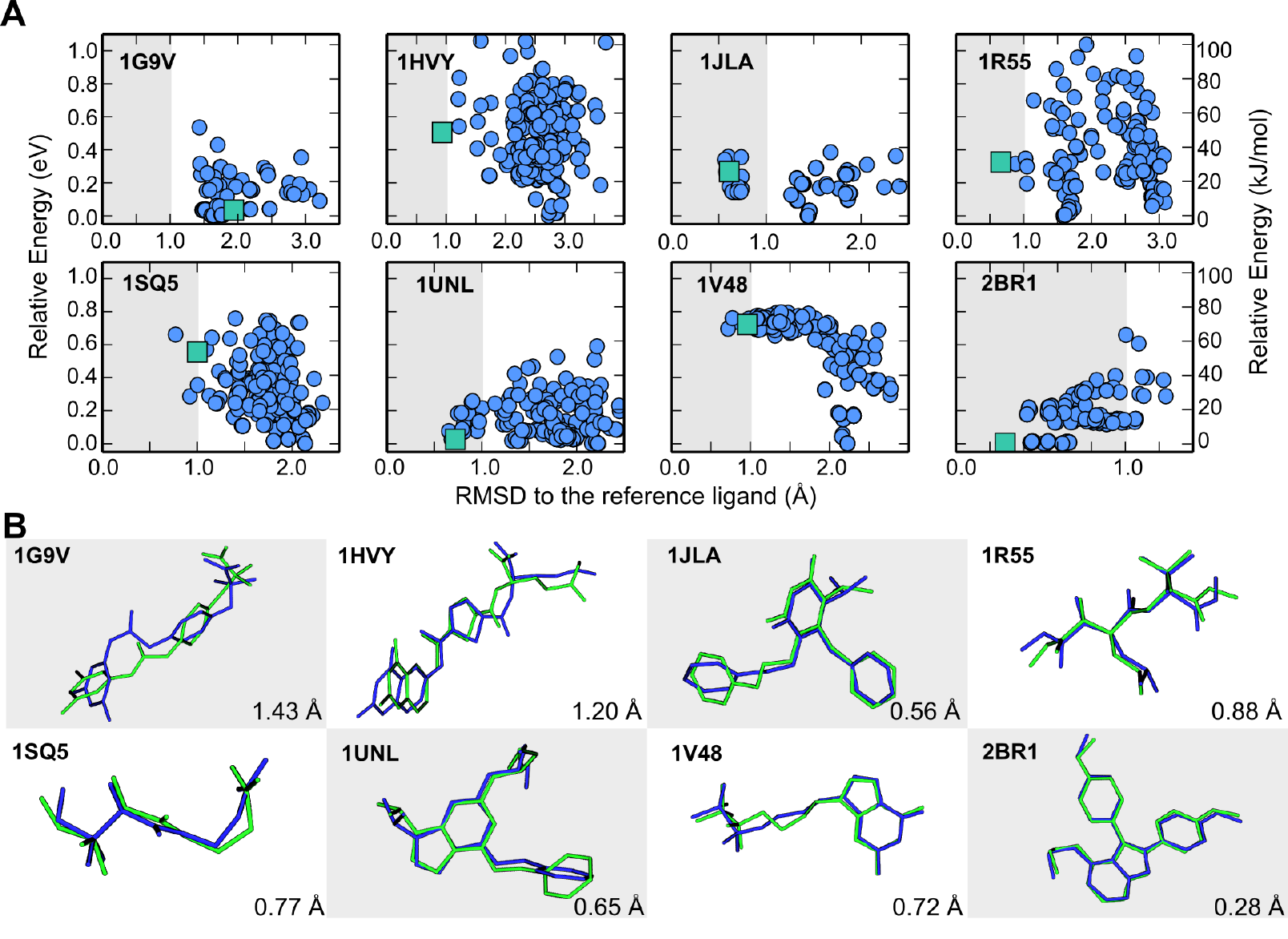}
 
  \caption{Evaluation of the results for the subset of the Astex Diverse Set. (A)~ Relative energy of all found conformers as a function of the RMSD to the reference ligand (blue circles). The green squares depict the reference ligand structures after DFT optimization. (B)~An overlay between the reference ligand (green) and the \textit{best match} (blue) is presented together with the corresponding RMSD value.}

  \label{astex_fig}
\end{figure}

We present the summary of the results in Table~\ref{tbl_astex_eval}. The number of found conformers is obtained after removing duplicates among all obtained conformers. For each of the molecules, we calculate the RMSD between the non-hydrogen atoms of each of the obtained structures and the reference ligand (Figure~\ref{astex_fig}A) (hydrogens in the Astex Diverse Set set are the result of modeling and not part of the experimental result). With this, we identify the \textit{best match}, i.e. the conformer which is most similar to the reference ligand. Furthermore, the reference ligand structures were optimized with DFT and are added to the respective plots for completeness. Figure~\ref{astex_fig}B shows the overlay between the reference ligand (before the DFT optimization) and the \textit{best match} for all molecules.

\begin{table}
  \caption{Selected ligands from the Astex Diverse Set. The number of found conformers is obtained after removing duplicates among all obtained conformers. The \textit{best match} is the conformer which is most similar to the reference ligand.}
  \label{tbl_astex_eval}
   \small
\begin{tabular}{|p{1.6cm}|p{1.6cm}|p{1.65cm}|p{1.75cm}|p{1.7cm}|p{1.7cm}|p{1.7cm}|p{1.7cm}|}
    \hline
     Target protein & No. of heavy atoms  & No. of rotatable bonds & No. of found conformers & \multicolumn{2}{|p{3.4cm}|}{RMSD (\r{A}) between the ligand and the} & \multicolumn{2}{|p{3.4cm}|}{$\Delta$E (eV) between the GA minimum and the}\\
     &&&& \textit{best match}  & GA minimum & \textit{best match} & optimized ligand\\
     \hline
     \hline
   1G9V & 25 & 8 & 70 & 1.43 & 1.66 &  0.536 & 0.035 \\    
   1HVY & 32 & 10 & 176 & 1.2 & 2.73 & 0.707&0.505\\   
   1JLA & 27 & 7 & 41 &  0.56&1.44 &  0.339&0.268\\  
   1R55 & 23 & 13 &  116 &0.88   & 1.59 & 0.315&0.326\\ 
   1SQ5 & 15 & 10 &  152 &0.77   & 2.14 &  0.661&0.555\\ 
   1UNL & 26 & 9 & 166 & 0.65 & 2.23 &  0.076&0.026\\  
   1V48 & 22 & 6 &  118& 0.72 & 2.23 & 0.696&0.722\\  
   2BR1 & 29 & 8 & 73 & 0.28 & 0.63 & 0.005&0.002\\

        \hline     
  \end{tabular}
\end{table}

For all investigated systems, we obtain a large number of conformers that are spread over a wide energy window. This satisfies our primary goal of obtaining a diverse set of conformers with a reliable energy hierarchy in a straightforward fashion. Moreover, in most of the cases, the RMSD between the \textit{best match} and reference ligand is satisfactory (i.e. below 1.0 \r{A}). Here we would like to note, that our energy evaluations are performed in the gas phase while the reference ligand is obtained from a X-ray crystal structure. The energy of the \textit{best match} is significantly higher than the energy of the GA minimum in  most of the cases. This finding supports the need for providing a broad range of conformers instead of only focusing on the global minimum of the particular energy function.

A few cases require further analysis. For two ligands, with the targets 1G9V and 1HVY, the RMSD values between the \textit{best match} and the reference ligand structure exceed the threshold value (1.0 \r{A}). One possible reason is the fact, that the reference ligand is not a minimum on the PES sampled by the GA. Another trivial cause might be the insufficient number of the performed GA runs.

Further we note that the optimization of the orientation of the hydroxy groups is required for obtaining a meaningful conformational ensemble. 

Apart from performing short GA runs, one long GA run has been performed for each of the selected molecules for comparison. In order to obtain comparable results (by means of the number of performed DFT optimization), following parameters have been adjusted: \textbf{max\_iter}=80, \textbf{iter\_limit\_conv}=70. We compare the results in terms of: (i) energy of the most stable structure, (ii) matching the reference ligand and (iii) number of found conformers. Detailed data about the results of the single long runs are given in Table~\ref{tbl_astex_long}. In terms of finding the \textit{best match}, the long GA run performs better than the three short GA runs together for some of the molecules. On the other hand, the number of found structures is significantly higher if three short GA runs are performed instead of a single long run for most of the molecules.

The results of the exploratory structure searches for the 8 drug-like ligands suggest that performing one long GA run instead of 3 short GA runs may increase the chance for finding the global minimum and simultaneously decrease the number of identified unique conformers.

\begin{table}
  \caption{Comparison of the results obtained from one long GA run (max. 80 iterations) and three short GA runs (each max. 30 iterations). $\Delta$E is the difference between the most stable structures found in the compared setups. }
  \label{tbl_astex_long}
   \small
\begin{tabular}{|p{2.2cm}|p{2.2cm}|p{2.2cm}|p{2.2cm}|p{2.2cm}|p{2.2cm}|}
    \hline
     Molecule & \multicolumn{2}{|p{4.4cm}|}{Number of found conformers} & \multicolumn{2}{|p{4.4cm}|}{RMSD (\r{A}) between the ligand and the \textit{best match}} &$\Delta$E (eV)  \\
     
     & 1 x max. 80 iterations & 3 x max. 30 iterations & 1 x max. 80 iterations & 3 x max. 30 iterations & \\
     \hline
     \hline
   1G9V & 45 & 70 & 0.57 & 1.43 & 0.0  \\    
   1HVY & 146 & 176 & 1.07 & 1.2 & 0.211 \\   
   1JLA & 40 & 41 & 0.62 & 0.56 &-0.005 \\    
   1R55 & 85 &  116 &0.69&  0.88   & 0.081\\ 
   1SQ5 & 99  &  152 &0.64& 0.77   & 0.031 \\ 
   1UNL & 93 & 166 &0.8&  0.65 & 0.003 \\  
   1V48 & 115 &  118& 0.81&0.72 & 0.054 \\  
   2BR1 & 47 & 73 & 0.28 &0.28 & -0.005 \\

        \hline     
  \end{tabular}
\end{table}



\newpage
\bibliography{first_principles_ga.bib}

\begin{mcitethebibliography}{87}
\bibitem[tor()]{torsionOH}
{We treat all X-X-O-H torsion angles as rotatable bonds}\relax
\mciteBstWouldAddEndPuncttrue
\mciteSetBstMidEndSepPunct{\mcitedefaultmidpunct}
{\mcitedefaultendpunct}{\mcitedefaultseppunct}\relax
\EndOfBibitem
\end{mcitethebibliography}

\end{document}